\documentclass[twoside,twocolumn,9pt]{article}
\usepackage{extsizes}
\usepackage[super,sort&compress,comma]{natbib}
\usepackage[version=3]{mhchem}
\usepackage{pdfpages} 
\usepackage[left=1.5cm, right=1.5cm, top=1.785cm, bottom=2.0cm]{geometry}
\usepackage{balance}
\usepackage{times,mathptmx}
\usepackage{sectsty}
\usepackage{graphicx}
\usepackage{lastpage}
\usepackage[format=plain,justification=justified,singlelinecheck=false,font={stretch=1.125,small,sf},labelfont=bf,labelsep=space]{caption}
\usepackage{float}
\usepackage{fancyhdr}
\usepackage{fnpos}
\usepackage[english]{babel}
\addto{\captionsenglish}{%
  
}
\usepackage{array}
\usepackage{droidsans}
\usepackage{charter}
\usepackage[T1]{fontenc}
\usepackage{setspace}
\usepackage[compact]{titlesec}

\usepackage{epstopdf}

\definecolor{cream}{RGB}{222,217,201}

\begin{document}

\pagestyle{fancy}
\thispagestyle{plain}
\fancypagestyle{plain}{
\renewcommand{\headrulewidth}{0pt}
}

\makeFNbottom
\makeatletter
\renewcommand\LARGE{\@setfontsize\LARGE{15pt}{17}}
\renewcommand\Large{\@setfontsize\Large{12pt}{14}}
\renewcommand\large{\@setfontsize\large{10pt}{12}}
\renewcommand\footnotesize{\@setfontsize\footnotesize{7pt}{10}}
\makeatother

\renewcommand{\thefootnote}{\fnsymbol{footnote}}
\renewcommand\footnoterule{\vspace*{1pt}%
\color{cream}\hrule width 3.5in height 0.4pt \color{black}\vspace*{5pt}}
\setcounter{secnumdepth}{5}

\makeatletter
\renewcommand\@biblabel[1]{#1}
\renewcommand\@makefntext[1]%
{\noindent\makebox[0pt][r]{\@thefnmark\,}#1}
\makeatother
\renewcommand{\figurename}{\small{Fig.}~}
\sectionfont{\sffamily\Large}
\subsectionfont{\normalsize}
\subsubsectionfont{\bf}
\setstretch{1.125} 
\setlength{\skip\footins}{0.8cm}
\setlength{\footnotesep}{0.25cm}
\setlength{\jot}{10pt}
\titlespacing*{\section}{0pt}{4pt}{4pt}
\titlespacing*{\subsection}{0pt}{15pt}{1pt}

\fancyfoot{}
\fancyfoot[RO]{\footnotesize{\sffamily{1--\pageref{LastPage} ~\textbar  \hspace{2pt}\thepage}}}
\fancyfoot[LE]{\footnotesize{\sffamily{\thepage~\textbar\hspace{3.45cm} 1--\pageref{LastPage}}}}
\fancyhead{}
\renewcommand{\headrulewidth}{0pt}
\renewcommand{\footrulewidth}{0pt}
\setlength{\arrayrulewidth}{1pt}
\setlength{\columnsep}{6.5mm}
\setlength\bibsep{1pt}

\makeatletter
\newlength{\figrulesep}
\setlength{\figrulesep}{0.5\textfloatsep}

\newcommand{\topfigrule}{\vspace*{-1pt}%
\noindent{\color{cream}\rule[-\figrulesep]{\columnwidth}{1.5pt}} }

\newcommand{\botfigrule}{\vspace*{-2pt}%
\noindent{\color{cream}\rule[\figrulesep]{\columnwidth}{1.5pt}} }

\newcommand{\dblfigrule}{\vspace*{-1pt}%
\noindent{\color{cream}\rule[-\figrulesep]{\textwidth}{1.5pt}} }

\makeatother

\twocolumn[
\begin{@twocolumnfalse}
  
\noindent\LARGE{\textbf{\ce{Ni80Fe20} Nanotubes with Optimized Spintronic Functionalities Prepared by Atomic Layer Deposition}} \\
\\

\noindent\large{Maria Carmen Giordano,\textit{$^{a}$} Simon Escobar Steinvall,\textit{$^{b}$} Sho Watanabe,\textit{$^{a}$} Anna Fontcuberta i Morral,\textit{$^{b,c}$} and Dirk Grundler$^{\ast}$\textit{$^{a,d}$}} \\
\\

\noindent\normalsize{\textbf{Abstract:} Permalloy \ce{Ni80Fe20} is one of the key magnetic materials in the field of magnonics. Its potential would be further unveiled if it could be deposited in three dimensional (3D) architectures of sizes down to the nanometer. Atomic Layer Deposition, ALD, is the technique of choice for covering arbitrary shapes with homogeneous thin films. Early successes with ferromagnetic materials include nickel and cobalt. Still, challenges in depositing ferromagnetic alloys reside in the synthesis via decomposing the consituent elements at the same temperature and homogeneously. We report plasma-enhanced ALD to prepare permalloy \ce{Ni80Fe20} thin films and nanotubes using nickelocene and iron(III) tert-butoxide as metal precursors, water as the oxidant agent and an in-cycle plasma enhanced reduction step with hydrogen. We have optimized the ALD cycle in terms of Ni:Fe atomic ratio and functional properties. We obtained a Gilbert damping of 0.013, a resistivity of 28 $\mu\Omega$cm and an anisotropic magnetoresistance effect of 5.6~$\%$ in the planar thin film geometry.  We demonstrate that the process also works for covering GaAs nanowires, resulting in permalloy nanotubes with high aspect ratios and diameters of about 150 nm. Individual nanotubes were investigated in terms of crystal phase, composition and spin-dynamic response by microfocused Brillouin Light Scattering. Our results enable NiFe-based 3D spintronics and magnonic devices in curved and complex topology operated in the GHz frequency regime.} \\
\vspace{1cm}
\end{@twocolumnfalse}

]

\renewcommand*\rmdefault{bch}\normalfont\upshape
\rmfamily
\section*{}
\vspace{-1cm}


\footnotetext{\textit{$^{a}$~Institute of Materials, Laboratory of Nanoscale Magnetic Materials and Magnonics, Ecole Polytechnique Federale de Lausanne, School of Engineering, 1015 Lausanne, Switzerland. E-mail: dirk.grundler@epfl.ch}}
\footnotetext{\textit{$^{b}$~Institute of Materials, Laboratory of Semiconductor Materials, Ecole Polytechnique Federale de Lausanne, School of Engineering,1015 Lausanne, Switzerland }}
\footnotetext{\textit{$^{c}$~Institute of Physics, School of Natural Sciences, Ecole Polytechnique Federale de Lausanne, 1015 Lausanne, Switzerland }}
\footnotetext{\textit{$^{d}$~Institute of Microengineering, School of Engineering, Ecole Polytechnique Federale de Lausanne, 1015 Lausanne, Switzerland }}






\section{Introduction}
Magnetic thin films play an extremely important role in technologies such as data storage media\cite{KRYDER1992}, GMR sensors \cite{Lenz2006}, spin valves\cite{Kools96} and  magnetic MEMS \cite{NIARCHOS2003, NiFeMEMS2019}. Among various magnetic materials, the NiFe alloy with stoichiometry \ce{Ni80Fe20}, often referred as permalloy, is technologically relevant for the high permeability, low coercivity, near zero magnetostriction and significant anisotropic magnetoresistance (AMR).
Permalloy plays a crucial role also in fundamental and applied research, as it is one of the standard materials chosen for the study of novel structures in nanomagnetism and magnonics\cite{Parkin2008, Neusser2009}. The latter is a modern branch of magnetism focused on the study of the spin waves, the collective excitations of magnetically ordered materials. Low energy consumption of spin waves and the potential compatibility with next-generation circuits beyond CMOS electronics make them a potential tool for non-charge based signal processing, communication and computation \cite{Neusser2009, Kruglyak2010, Nikitov_2015, Stancil2009}. In magnonics, the choice of magnetic materials would fall on those with low damping for spin waves, a property quantified by the phenomenological dimensionless Gilbert damping parameter $\alpha$. Standard materials would be insulating magnets as yttrium iron garnet (YIG) and polycristaline metallic alloys as NiFe and CoFeB. While YIG offers the lowest spin wave damping, the choice of polycristaline metallic alloys as NiFe and CoFeB meets better the needs for the industrial scalability of miniaturized and integrated systems\cite{Gubbiotti2019}.\\
So far, \ce{Ni80Fe20} films have been mostly obtained by physical vapor deposition (PVD) methods, like sputtering \cite{WANG2013405, Ohtake2012}, electron beam evaporation and molecular beam epitaxy\cite{Ohtake2011}. Other routes are electrodeposition\cite{Tabakovic2014,BALACHANDRAN2009,Medina2018,Bialostocka2018}, spray coating and micromolding \cite{Cortes2014}. The PVD techniques are well suited for the production of planar nanostructures\cite{Pancaldi2019, Bhat2020}. However, they are limited by shadowing effects, which make them disadvantageous for the coating of three-dimensional (3D) nanostructures with high aspect ratios. Correspondingly, tetrapods for multi-branched 3D spintronics were prepared from polycrystalline cobalt by means of electrodeposition \cite{Sahoo2018}. To further optimise the deposition of 3D permalloy coatings, it is a necessity to increase the step edge coverage. This enhances the deposition uniformity irrespective of the morphology of the deposited surface, and in turn opens up the study of new physical phenomena envisioned in 3D nanomagnetic systems \cite{Fischer2020, Gubbiotti2019, Otalora2017, Streubel2016, Yan2011, Fernandez-Pacheco2017, Parkin2008, Landeros2007, ESCRIG2007}.
Atomic layer deposition (ALD) is a chemical deposition technique offering the ideal conformality on 3D nanostructured surfaces. Here, the thin film formation is based on the repeated exposure of a substrate to separate precursors \cite{Johnson2014,Ozaveshe2019}. These precursors react with the surface of a material one at a time in a sequential, self-limiting manner ensuring a good control of thickness, excellent step coverage and conformality on substrates with different geometries and aspect ratios. The deposition of high-$\kappa$ gate oxides, such as \ce{Al2O3}, has been one of the most widely examined areas of ALD and already advances microelectronics applications like metal-oxide semiconductor field effect transistors (MOSFET) and dynamic random access memories (DRAM)\cite{Ritala2009}. The processes for depositing metal layers using this technique, including ferromagnetic metals, still present technical challenges and are the subject of extensive research \cite{Lim2003, Wang2016, Klesko2016, Kerrigan2018, Hagen2019}. The main challenges are the limited number of suitable precursors, the difficulty in reducing metal cations and the tendency of metals to agglomerate into islands \cite{Hagen2019,Lim2003}. We emphasize the difficulty, specific to a bi-metallic alloy, to identify a pair of precursors, one for each metal, which have a similar reactivity with the substrate surface, with a suitable co-reactant and in the same thermal range. Many works concern the use of ALD to obtain 3D nickel nanostructures such as nanotubes \cite{Daub2007,Ruffer2012,Buchter2013,Weber2012,Ruffer2014,Huber}. The research on NiFe metallic alloys deposited by ALD is still in its infancy.  Ferrimagnetic oxides Ni$_{x}$Fe$_{3-x}O_{4}$ have been successfully prepared by exploiting the ALD process of the two binary oxides NiO and FeO \cite{Chong2010}. In the work of Espejo $et al.$ \cite{Espejo2016} a study was presented where a combination of supercyclic ALD with thermal reduction is used to achieve metallic alloy NiFe thin films. The process explored was based on nickelocene and ferrocene as Ni and Fe precursors, ozone as co-reactant to get an intermediate NiFe oxide and a further annealing in hydrogen. A linear dependency between the Ni : Fe precursors pulse ratio and the final stoichiometry was identified, leading to the achievement of a NiFe stoichiometry (\ce{Ni83Fe17}) very close to that of permalloy.  The study showed how the initial amount of oxygen in the NiFe - oxide thin films obtained induces a significant dewetting phenomenon during the thermal reduction in hydrogen, compromising the morphology in the final metallic form.\\
In this work we present a different approach where the nickel (iron) sequence exploits nickelocene (iron(III) tert-butoxide) as metal precursors, water as the oxidant agent and an in-cycle plasma enhanced reduction step with hydrogen. Ni-rich NiFe thin films with different Ni:Fe atomic ratios were achieved by alternating $m$ times the sequence for the deposition of nickel \cite{Giordano2020, Chae2002} with a single sequence for the deposition of oxidized iron \cite{Bachmann2007, Escrig2008, Bachmann2009} and a post-deposition annealing treatment in hydrogen.
The planar thin films so prepared were characterized in terms of a series of electric and magnetic properties considered significant for industrial applications and research. We quantitatively compare them with thin films that we prepared by the electron-beam evaporation using a commercial \ce{Ni80Fe20} target material.
The sequence with optimized Ni:Fe pulse ratio $m$, combined with the post-processing annealing treatment, allowed us to prepare thin films with the expected \ce{Ni3Fe} FCC phase ($L1_{2}$) \cite{NiFe2012}, stoichiometry extremely close to target permalloy (Ni$_{80.4}$Fe$_{19.6}$) and physical properties unprecedently measured in permalloy thin films prepared by ALD. We measured a coercive field of 3.6 mT, a Gilbert damping of 0.013  and a resistivity of 28 $\mu\Omega$cm. The process presented could contain the dewetting phenomenon on silicon substrates. As a proof of concept for the use of this technique to obtain nanostructured 3D coatings, we show the fabrication of NiFe nanotubes (NTs) using GaAs nanowires (NWs) with high aspect ratio and diameters lower than 100 nm as nanotemplates. We provide a chemical and structural characterization of the NiFe nanotubes, confirmed to have a permalloy shell. Lastly, the low damping of the ferromagnetic shell allowed to measure several spin waves resonant modes in individual NTs investigated by micro-focused Brillouin Light Scattering ($\mu$-BLS). Our results show the remarkable potential of the ALD technique to venture the third dimension in nanomagnetism, magnonics and spintronics applications based on thin-film permalloy prepared by conformal coating technique.

\section{Results and discussion}

\subsection{Morphology of Nickel-Iron Thin Films and Nanotubes}

\begin{figure*}[h]
\centering
	\includegraphics[width=0.9\textwidth]{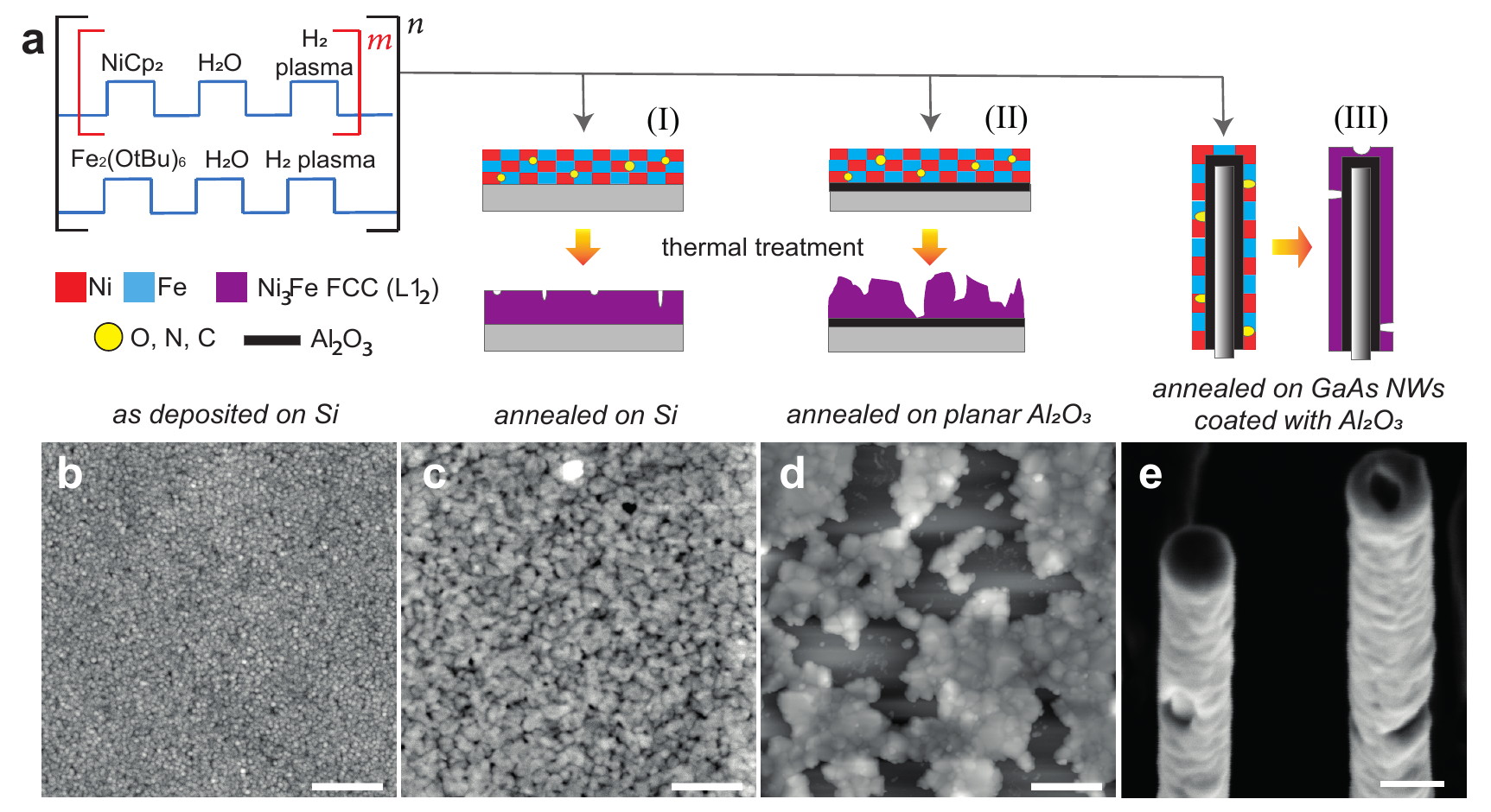}
	\caption{(a) Sketch describing the PEALD cycle for the deposition of Ni$_{100-x}$Fe$_{x}$ originally contaminated by residual process elements (O, N, C); the annealing treatment to induce the \ce{Ni3Fe} FCC phase ($L1_{2}$) \cite{NiFe2012} formation and the observed morphology of the final thin films achieved, on different substrates. Atomic force microscopy performed on a $2~\mu$m$\times 2~\mu$m area of the sample prepared with samples with $m=6$ (b) as-deposited, (c) annealed on a  bare Si substrate (color code range: -7 nm $\div$ +7 nm; scale bar: 400 nm), (d) annealed on a Si substrate pre-coated by 5 nm of \ce{Al2O3} (color code range: -70 nm $\div$ +70 nm;scale bar: 400 nm). (e) SEM micrograph of Ni$_{100-x}$Fe$_{x}$ nantoubes on GaAs nanowires pre-coated by 5 nm of \ce{Al2O3}, after the annealing treatment (scale bar: 100 nm) }\label{Figure 1}
\end{figure*}

In Figure~\ref{Figure 1}a we show schematically the process followed to prepare Ni$_{100-x}$Fe$_{x}$ thin films by plasma enhanced atomic layer deposition (PEALD) on different substrates. The PEALD cycle combines a sequence for the deposition of nickel\cite{Chae2002, Giordano2020} and a sequence for the deposition of \ce{Fe2O3}\cite{Bachmann2007, Escrig2008, Bachmann2009}, to which we have added a plasma hydrogen step. The Ni:Fe atomic ratio is controlled by the number $m$ of the Ni sequence repetition between each Fe step.
The NiFe process was tested on relevant substrates like planar Si wafers, both uncoated and coated with few nm of alumina \ce{Al2O3}, and GaAs NWs encapsulated in a capping layer of \ce{Al2O3}. The as-deposited films were annealed aiming at the \ce{Ni3Fe} FCC phase.  We show the AFM analysis performed on the sample prepared with $m=6$ in its as-deposited (Figure~\ref{Figure 1}b) and annealed state (Figure~\ref{Figure 1}c) on a bare Si substrate. In Figure~\ref{Figure 1}d we report the AFM analysis on the annealed thin film obtained with the same process, exploiting $m=6$, on a Si substrate coated with 5 nm of alumina. Lastly, the SEM micrograph in Figure~\ref{Figure 1}e depicts the morphology of the thin films deposited on GaAs/\ce{Al2O3} nanowires, once the annealing treatment is performed. Depending on the type of substrate, the thermal treatment was found to have different effects on the final morphology of the thin films. In general, on a bare Si substrate, we observe that the annealing process is accompanied by a thin film thickness shrinkage of about 15 $\%$ , the formation of nanopores and a higher surface roughness with respect to the as-deposited thin film. For the reported sample the rms roughness increased from 1.2 nm to 2.1 nm in thin films with a thickness of 27 nm and 23 nm, respectively, in the as-deposited (Figure~\ref{Figure 1}b) and annealed form (Figure~\ref{Figure 1}b). For the Si substrate coated with alumina we observe that the thermal treatment is accompanied by a dewetting of the thin film. Here, the thin film arranges in agglomerates to reduce the film-substrate interface area and minimize the system energy. In Fig.~\ref{Figure 1}e, we observe the formation of random localized small holes in the 3D shells achieved by coating GaAs/\ce{Al2O3} NWs.  Despite the unfavorable adhesion of Ni$_{100-x}$Fe$_{x}$ on the alumina layer, the encapsulation of GaAs nanowires by few nm of this material was necessary to prevent the As evaporation during the annealing and to use them as a template for  Ni$_{100-x}$Fe$_{x}$ nanotubes.\\
The same experiments were done with Ni : Fe pulse ratio 4 $\leq m \leq 7$. The thermal treatment was performed for all the compositions. All the samples were amorphous and contained remaining oxygen, nitrogen and carbon in their as-deposited state (Figure~S1). The types of morphology described are common to the samples obtained with the synthesis parameter m = 4, 5 and 6. For the sample prepared with m = 7, more nanoholes are observed (Figure~S2). For this sample, we speculate that phase segregation might have further increased its roughness and the number of holes during the annealing treatment and subsequent cooling, as its composition might be close to the \ce{Ni90Fe10} phase segregation line in the Ni-Fe phase diagram \cite{NiFe2012}.

\subsection{Growth rates}
The choice of the homoleptic dinuclear iron(III) tert-butoxide complex (\ce{Fe2(OtBu)6}) as iron precursor is motivated by its capability to react with water in a self-limiting manner. Furthermore, it reacts at temperatures ($T = 130 - 180~^\circ$C \cite{Bachmann2007, Escrig2008, Bachmann2009}) compatible with the one for the deposition of nickel using nickelocene (\ce{NiCp2}), water and plasma hydrogen as reactants ($T = 170~^\circ$C in our system \cite{Giordano2020}). The growth rate of the processes exploiting $m = 4, 5, 6$ and 7 was estimated to be (0.15, 0.17, 0.18 and 0.20) \AA/ cycle, respectively. The recorded values were lower than those that would be obtained for the Ni and \ce{Fe2O3} processes individually. We hypothesize that the deposition of Ni in the presence of \ce{Fe2O3} is slower than in the presence of Ni alone and that a partial reduction of \ce{Fe2O3} in the presence of Ni and plasma hydrogen during the deposition sequence occurs. The in-depth study of these co-deposition mechanisms is not the subject of this work.\\

\subsection{Stoichiometry and Crystallographic Analyisis of Ni$_{100-x}$Fe$_{x}$ Thin Films}
\begin{figure}[h]
	\includegraphics[width=0.5\textwidth]{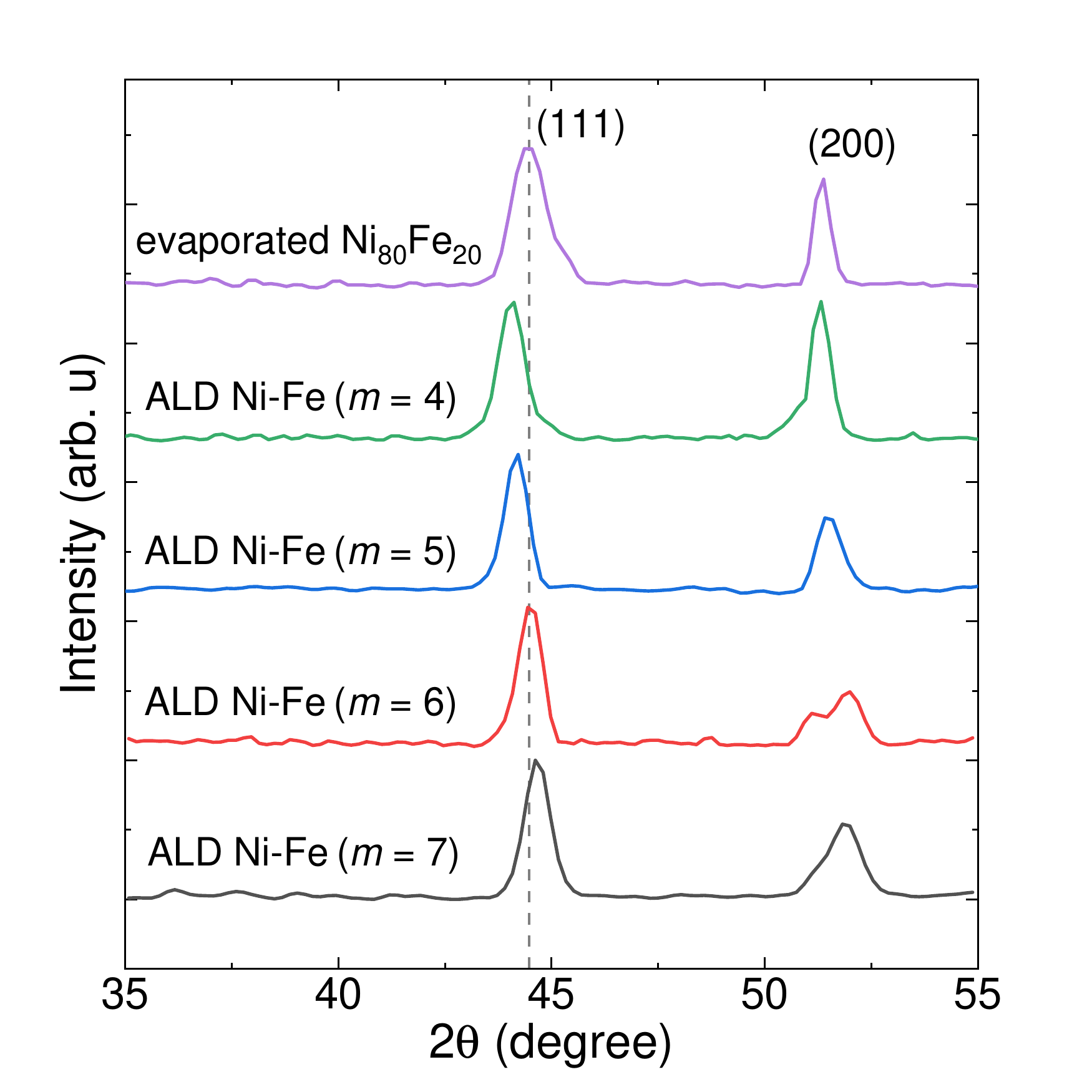}
	\caption{X-ray diffractograms of 50 nm thick evaporated commercial \ce{Ni80Fe20} thin film and of $22~\pm3$ nm thick Ni$_{100-x}$Fe$_{x}$ films prepared by PEALD with $m$ ranging from 4 to 7 (see labels). The position of the (111) reflection of the evaporated \ce{Ni80Fe20} is marked by a dotted line. The ALD thin films were deposited at 170 $^\circ$C on a silicon substrate and annealed at 380$~^\circ$C.}\label{Figure 2}
\end{figure}

Figure~\ref{Figure 2} shows the X-ray diffractograms (XRD) of the annealed planar thin films prepared by PEALD with $ m=4,5,6$ and $7$, compared with the diffractogram of an electron-beam evaporated thin film of commercial permalloy \ce{Ni80Fe20}, displayed at the top of the graph.
The XRD results show mainly two peaks compatible with the (111) and (200) reflections of the \ce{Ni3Fe} FCC phase (ICSD code: 632930). The gray dotted line marks the position of the (111) reflection of the permalloy thin film evaporated from commercial target material. The (111) peak linewidth and position are reported in Table~\ref{Table 1} for each sample. The small linewidth of the ALD thin films, compared to the reference permalloy sample, reflects a larger average crystallite size. We attribute this observation to the higher temperatures experienced both during the deposition and the annealing process by the samples prepared by ALD. The shift towards a lower angle of the (111) peak, going from the sample prepared with $m=7$ to the one prepared with $m=4$, reflects the lattice expansion caused by the incorporation of a higher atomic percentage of Fe.
The EDS analysis on the samples (Table~\ref{Table 1}) confirms this hypothesis.
\begin{table*}[h]
\centering
  \caption{Stoichiometry and Crystallographic Information Extracted from the Energy Dispersive Spectra and the X-Ray Diffractograms of the Annealed Ni$_{100-x}$Fe$_{x}$ Thin Films as a Function of the Ni:Fe Pulse Ratio $m$.}
  \label{Table 1}
  \begin{tabular*}{0.9\textwidth}{@{\extracolsep{\fill}}ccccccc}
    \hline
Ni:Fe pulse ratio & Ni (at$\% $) & Fe (at$\% $)& Phase & (111) peak position & $FWHM_{(111) peak}$ & Lattice parameter \\
 $m$ &  & $x$ &  & ($^\circ 2\theta$) & ($^\circ 2\theta$) & $a$(\AA)\\
    \hline
     4 & 76 $\pm$ 5 & 24 $\pm$ 5 & FCC ($L1_{2}$) & 44.063 & 0.634 & 3.557\\
     5 & 78 $\pm$ 5 & 22 $\pm$ 5 & FCC ($L1_{2}$) & 44.167 & 0.543 & 3.549\\
     6 & 81 $\pm$ 5 & 19 $\pm$ 5 & FCC ($L1_{2}$) & 44.493 & 0.544 & 3.524\\
     7 & 83 $\pm$ 5 & 17 $\pm$ 5 & FCC ($L1_{2}$) & 44.666 & 0.615 & 3.511\\
     \hline
     Evaporated \ce{Ni80Fe20} & 80 $\pm$ 5 & 20 $\pm$ 5 & FCC ($L1_{2}$) & 44.666 & 0.802  &3.525\\
    \hline
  \end{tabular*}
\end{table*}

Remarkably, the Fe incorporation in the lattice is accompanied by the appearance of stress in the film, which involves strains of the (200) plane. This is particularly visible in the diffractogram of the sample prepared with $m=6$ where we can distinguish two components of the (200) reflection, possibly indicating that the stress relaxation was achieved by the formation of two domains with different (200) spacing. The lattice parameter of the samples was determined based on the (111) position (Table~\ref{Table 1}). The lattice parameter is found to scale linearly with the iron content in this composition range.
The lattice constant of the permalloy thin film reference, as well as the Ni:Fe atomic ratio measured by the SEM-EDS analysis, match closely the parameters obtained for the ALD sequence with Ni:Fe pulse ratio $m=6$. For this pulse ratio in particular we verified the stoichiometry and the phase of the Ni$_{100-x}$Fe$_{x}$ thin films obtained in the shape of nanotubes, with higher resolution techniques.
\begin{figure*}[h]
\centering
	\includegraphics[width=0.9\textwidth]{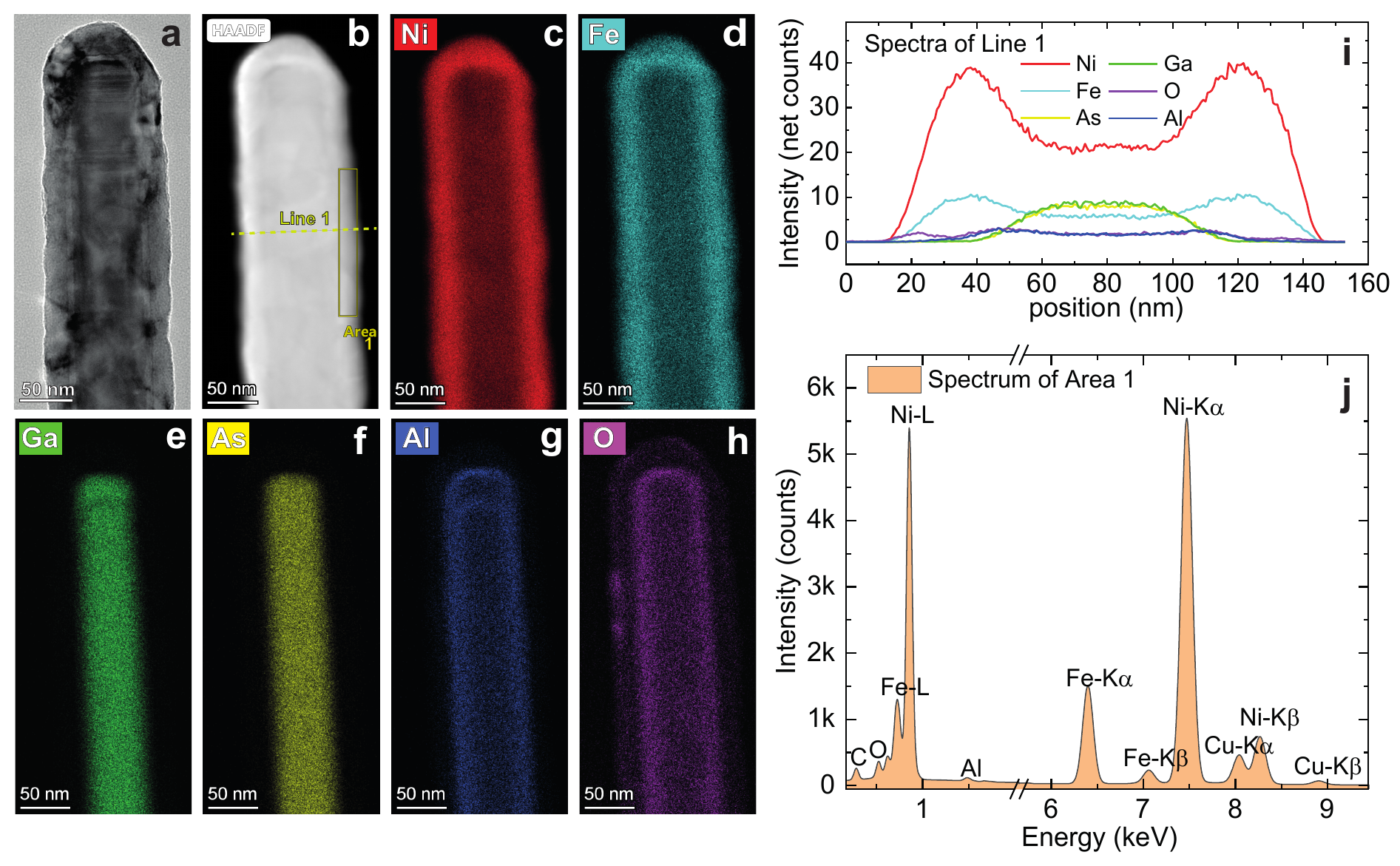}
	\caption{(a)BF-TEM image and (b) HADDAF image of the extremity of a Ni$_{100-x}$Fe$_{x}$ / \ce{Al2O3} nanotube prepared by ALD on a GaAs nanowire. STEM-EDX maps of (c,d) the Ni$_{100-x}$Fe$_{x}$ outer shell, (e,f) the GaAs nanowire core and (g,h) the \ce{Al2O3} spacing layer (scale bars: 50 nm). The HADDAF image applies to all the STEM-EDX maps. (i) Elements distribution along the Line 1 and (j) elemental analysis of the Area 1 depicted in (b). The outer Ni$_{100-x}$Fe$_{x}$ thin film were achieved with a deposition at 170 $^\circ$C, using a Ni:Fe pulse ratio $m=6$, and a further annealing treatment at 380$~^\circ$C.}\label{Figure 3}
\end{figure*}

In Figure~\ref{Figure 3} (and Figure~S3) we report on the elemental analysis by STEM-EDX of NiFe nanotubes deposited with the Ni:Fe precursor pulse ratio $m=6$. In particular, Figure~\ref{Figure 3}a and Figure~\ref{Figure 3}b show a bright field TEM image and the HADDAF image of the nanotube, respectively.  In Figure~\ref{Figure 3}c-h the element distribution of Ni, Fe, Ga, As, Al and O are reported. In Figure~\ref{Figure 3}i and Figure~\ref{Figure 3}j we report the elements distribution along the Line 1 and the EDX spectra of the Area 1 depicted in Figure~\ref{Figure 3}b , respectively. The NT images present a very smooth Ni-Fe shell, with no visible holes on a length of 350 nm. The core remained pristine despite the high-temperature annealing conditions. The elemental maps show a homogeneous intermixing of Ni and Fe in the outer shell (Figure~\ref{Figure 3}c,d), a GaAs nanowire core (Figure~\ref{Figure 3}e,f) and an intermediate alumina capping layer (Figure~\ref{Figure 3}g,h). Apart from the expected presence of oxygen in the \ce{Al2O3} layer at the interface between the GaAs core and Ni-Fe shell, the oxygen is almost absent in the outer shell, indicating that the thermal treatment in hydrogen successfully reduced the oxidized metal atom (Figure~\ref{Figure 3}h).  The elemental distribution along Line 1 shows at which position, in the 3D object, we measure the higher counts of each element. This analysis confirms that the elements of the three layers did not intermix with each other. The chemical analysis of Area 1 gives the following composition: 77.4 at$\%$ Ni, 19.1 at$\%$ Fe, 3.5 at$\%$ O. The final NT material has, therefore, a Ni:Fe atomic ratio equal to 80.2 : 19.8 and a content of oxygen equal to 3.5 at$\%$. For this batch of NTs, we estimated an average stoichiometry of Ni$_{80.4 \pm 0.3}$Fe$_{19.6\pm 0.3 }$, based on the analysis of five elements (Figure~S3, Table~S1). The amount of oxygen within the shell of the NTs has been quantified as ($3.5~\pm0.1$) at$\%$ and is compatible with a small superficial passivation of the shell.
After having verified the NTs stoichiometry, we analyzed the crystal phase of the 3D coatings by high resolution transmission electron microscopy (HR-TEM).
\begin{figure*}[h]
\centering
	\includegraphics[width=0.9\textwidth]{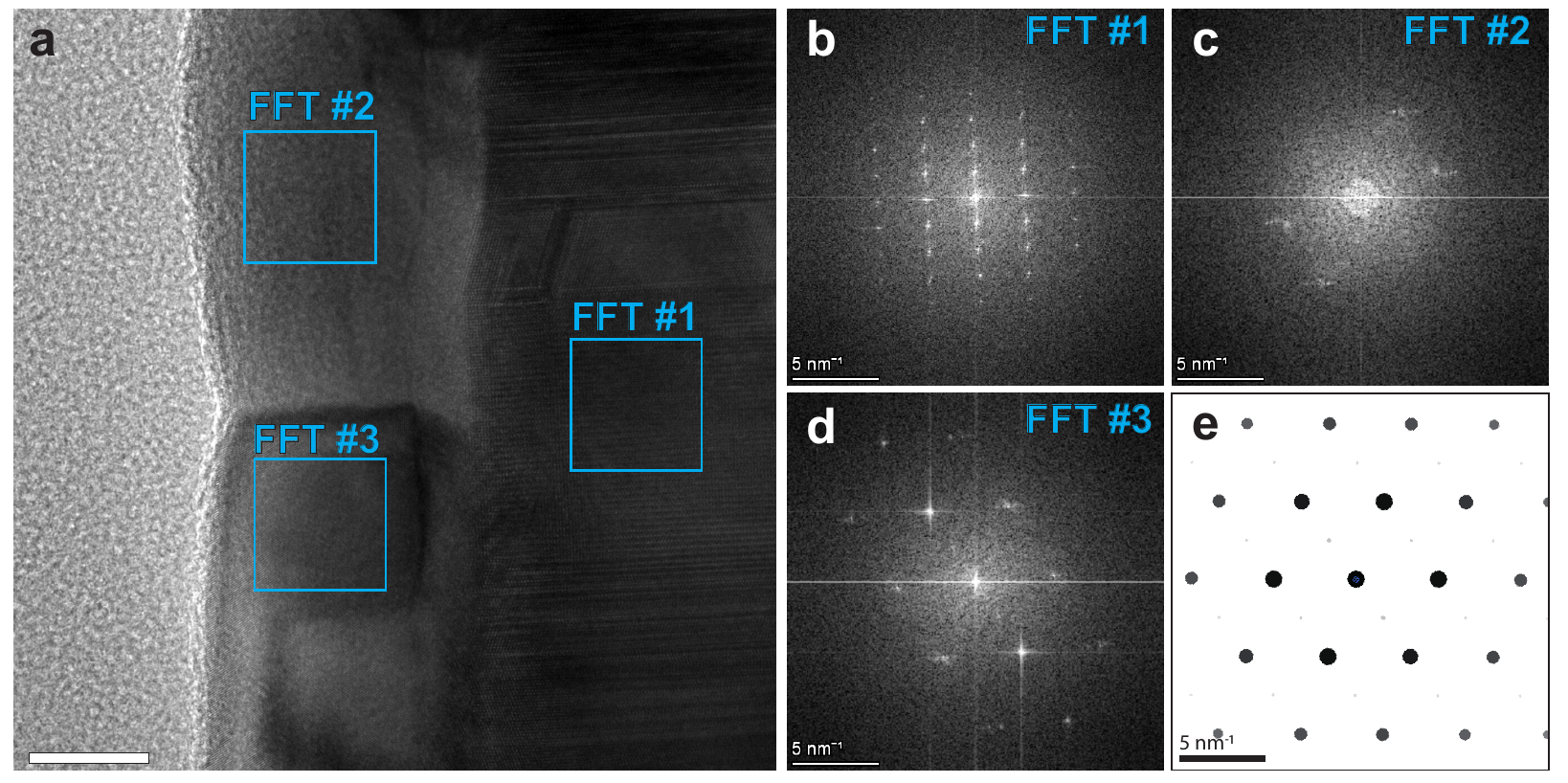}
	\caption{(a) HR-TEM image showing the single-crystal GaAs core, the \ce{Al2O3} spacer and the nanocrystalline \ce{Ni80Fe20} outer shell (scale bar: 10 nm). Power spectra were extracted for the core (b) and the outer shell (c-d). (e) Simulated diffraction patterns for the \ce{Ni3Fe} FCC phase along the [101] zone axis, exhibiting the same symmetry as (d). }\label{Figure 4}
\end{figure*}

Figure~\ref{Figure 4}a shows an HR-TEM image of the interface between the NiFe / \ce{Al2O3} nanotube, prepared by ALD, and the GaAs nanowire core. The NiFe shell results locally compact and smooth. We took the fast Fourier transform (FFT) of the HR-TEM image in correspondance of the regions highlighted by blue squares. The extracted power spectra of the spatial frequencies are reported in Figure~\ref{Figure 4} b-d, and show the single-crystal GaAs core (Fig.~\ref{Figure 4}b), and the nanocrystalline \ce{Ni80Fe20} shell (Fig.~\ref{Figure 4}c,d). The observed pattern, and thus the orientation, varies between the \ce{Ni80Fe20} grains in the shell, indicative of a random crystal orientation. The symmetry of the pattern from the bottom grain (Fig.~\ref{Figure 4}d) can be matched to a simulated SAED pattern of \ce{Ni3Fe} (Fig.~\ref{Figure 4}e) along a [101] zone axis. This shows that the nanotube growth occurs in a randomly oriented nanocrystalline fashion with the standard \ce{Ni3Fe} FCC permalloy crystal structure ($L1_{2}$).

\subsection{Magnetic Properties of Ni$_{100-x}$Fe$_{x}$ Thin Films }
\begin{figure}[h]
\centering
	\includegraphics[width=0.45\textwidth]{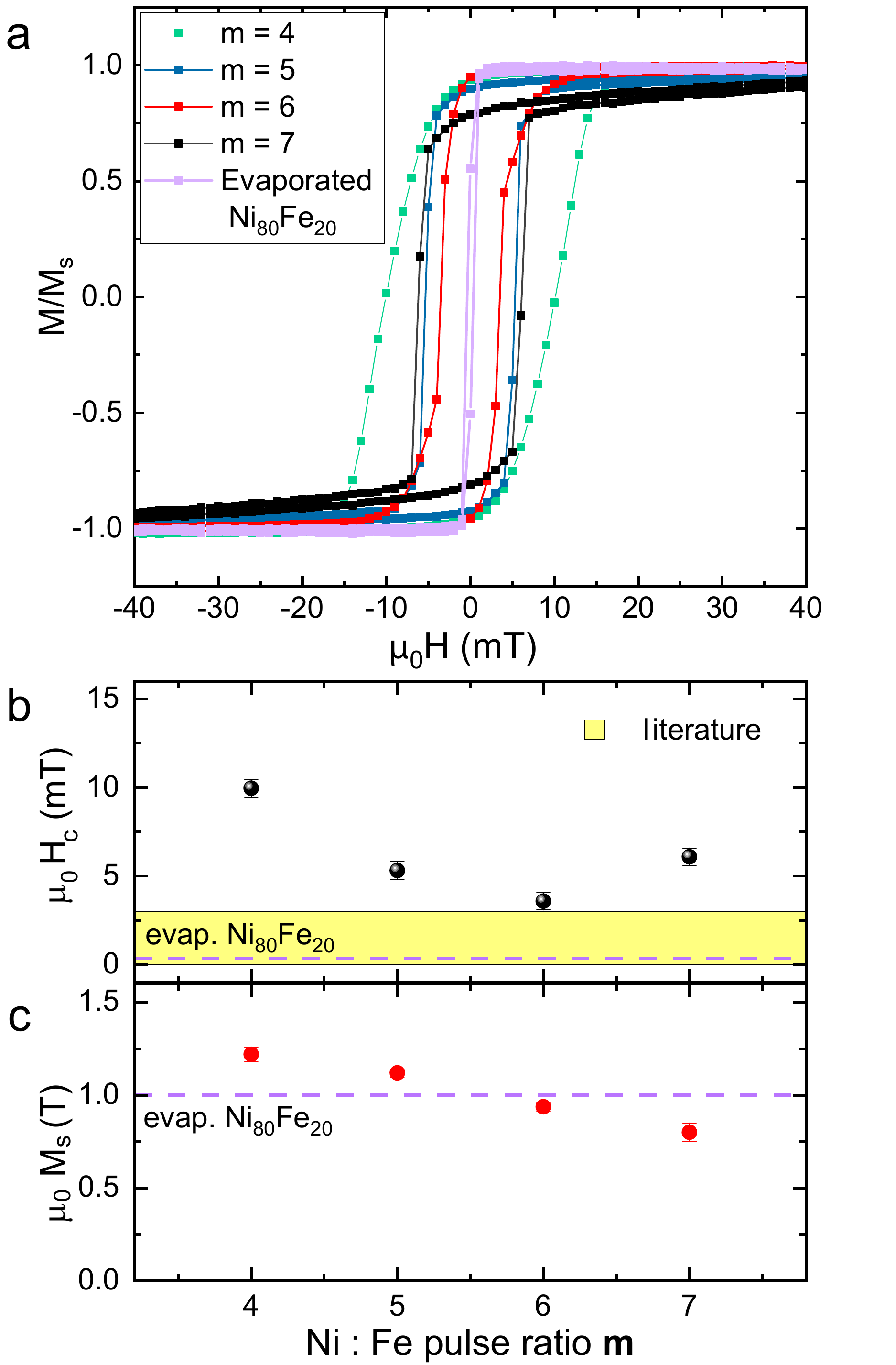}
	\caption{(a) Hysteresis loops measured by VSM at room temperature of annealed Ni$_{100-x}$Fe$_{x}$ thin films prepared by PEALD with Ni:Fe pulse ratios $m=4,5,6$ and 7 and of the reference \ce{Ni80Fe20} thin film. Values of (b) coercive fields $H_{c}$ and of (c) saturation magnetization $M_{\rm s}$ are plotted as a function of the ALD process parameter $m$. The literature values for permalloy $H_{c}$ are indicated with a yellow range, while the values of $H_{c}$ and $M_{\rm s}$ measured for the evaporated permalloy, prepared as a reference are displayed with a dashed line.}\label{Figure 5}
\end{figure}

The magnetic hysteresis of the as-deposited and annealed planar Ni$_{100-x}$Fe$_{x}$ thin films prepared by PEALD was acquired at room temperature using a vibrating-sample magnetometer (VSM). The ALD thin films show a weak magnetization in their as-deposited state (Figure~S4). Here we focus on the hysteresis of the annealed thin films prepared with Ni:Fe pulse ratios $ m=4,5,6$ and $7$ and compare the results with the hysteresis of the reference \ce{Ni80Fe20} thin film. Figure~\ref{Figure 5}a (Figure~S5) shows the hysteresis obtained with the field applied in plane at zero (45) degrees with respect to the edge of a squared sample. In Figure~\ref{Figure 5}b we report the extrapolated values of coercive field as a function of the synthesis parameter $m$. As a reference we report, in yellow, the range of coercive field values reported for permalloy in literature \cite{Glaubitz2011,Gao1997} and, marked with a dotted line, the value measured of the reference \ce{Ni80Fe20} thin film. The four ALD-prepared samples were ferromagnetic at room temperature and exhibited a coercive field which stayed constant with the angle of the applied field (Figure~S5). A hysteresis squareness $M_{r}/M_{\rm s}$ above 0.9 for samples prepared with $m=4,5$ and 6 and equal to 0.8 for the sample prepared with $m=7$ indicated that the magnetocrystalline anisotropy overall was small in the alloys \cite{NiFe_Mani}. The coercive fields measured for ALD-prepared Ni$_{100-x}$Fe$_{x}$ thin films ranged from 3.6 to 10 mT, with the minimum value being measured for the thin film prepared with Ni:Fe pulse ratio $m=6$. The coercive field of 3.6 mT was close to the values of 0 to 3~mT expected for permalloy \cite{Glaubitz2011,Gao1997} and one order of magnitude smaller than the value of 47.5 mT reported earlier for ALD-prepared Ni$_{100-x}$Fe$_{x}$ thin films \cite{Espejo2016}. The evaporated reference \ce{Ni80Fe20} thin film exhibited a coercive field of 0.36 mT. We attribute the discrepancy between the coercive fields of ALD-prepared Ni$_{100-x}$Fe$_{x}$ and evaporated commercial \ce{Ni80Fe20} thin films to the nanoholes induced by the annealing treatment of the ALD-prepared thin films. These defects might act, in fact, as domain wall pinning centers in the ALD-grown material, affecting the magnetization reversal.
\begin{figure}[h]
	\includegraphics[width=0.45\textwidth]{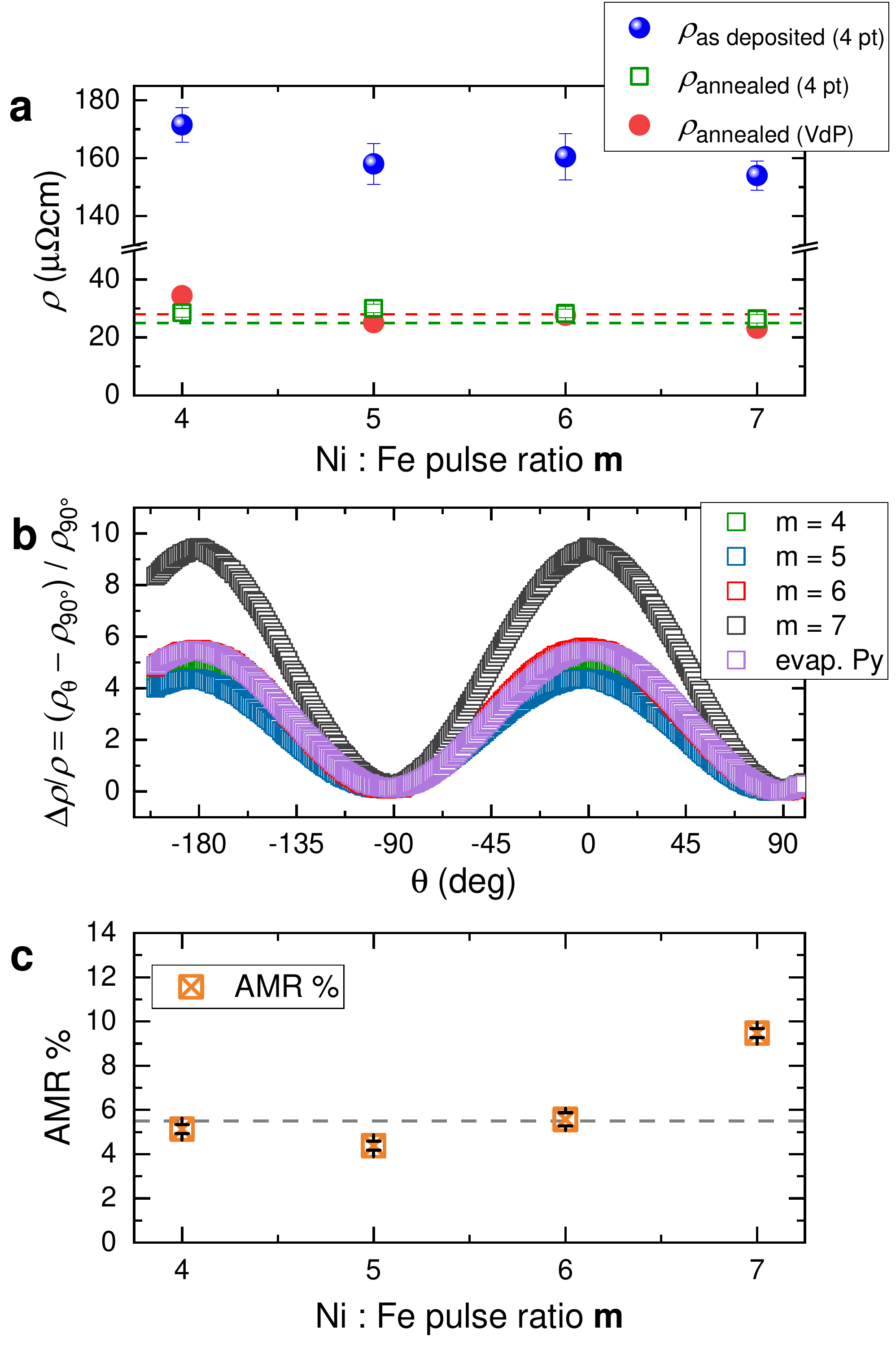}
	\caption{ (a) Resistivity values of the as-deposited and annealed ALD-prepared Ni$_{100-x}$Fe$_{x}$ thin films, compared with the value measured for a 20 nm evaporated permalloy thin film (dashed line). (b) Magnetoresistance measured at room temperature for a rotating in-plane field of 80 mT and (c)relative AMR effect extrapolated of the ALD-prepared Ni$_{100-x}$Fe$_{x}$ thin films and a 20nm - thick evaporated commercial \ce{Ni80Fe20} thin film used as a reference. }\label{Figure 6}
\end{figure}

In Figure~\ref{Figure 5}c, the saturation magnetization values $M_{\rm s}$ extracted from the hysteresis are shown as a function of $m$. The values of $\mu_{0}M_{\rm s}$ rang from 0.80 to 1.22 T and are found to scale linearly with the content of iron, i.e. $\mu_{0}M_{\rm s}$ decreases with the Ni:Fe pulse ratio $m$. The sample prepared with $m=6$ exhibits a saturation magnetization $\mu_0M_{\rm s}$ of 0.94 T, which is close to the value measured for permalloy (1T).

In Figure~\ref{Figure 6}a we show the measured values for the resistivity both in the as-deposited and in the annealed thin films, as a function of the parameter $m$. The resistivities measured for the annealed Ni$_{100-x}$Fe$_{x}$ thin films have been obtained with a commercial 4-point probe station and a custom-built Van der Pauw configuration measurement setup. The dashed lines correspond to the resistivity values measured on a 20 nm thin film of evaporated \ce{Ni80Fe20}. In Figure~\ref{Figure 6}b, we depict the magnetoresistance measured on planar NiFe thin films when an applied in-plane magnetic field $\mu_{0}H$ = 80 mT was rotated. The angle $\theta$ was defined between the current direction and $\textbf{H}$. We display the anisotropic magnetoresistance (AMR) as $ \dfrac{\Delta \rho(\theta)}{\rho} $ = $\dfrac{\rho(\theta) - \rho(90^{\circ})}{\rho(90 ^{\circ})}$. The relative AMR effect is reported in Figure~\ref{Figure 6}c as a function of the synthesis parameter $m$.
We obtain resistivity values ranging from 154 to 172 $\mu\Omega$cm for the as-deposited thin films. The resistivity is found to decrease with increasing $m$, hence with the nickel content. Our resistivity values are much lower when compared with those obtained in the process that used ozone as a co-reactant \cite{Espejo2016, Huber}. We attribute this observation to the reduced amount of oxygen in the as-deposited material, that we achieved by the specific choice of precursors and reactants. The values of resistivity range from 23 to 34 $\mu\Omega$cm in the annealed samples and do not vary significantly with $m$. At $m=6$ we get $\rho$ = 28 $\mu\Omega$cm. The values measured match closely the ones measured for the evaporated commercial \ce{Ni80Fe20}, marked by the dashed lines. Hence, we speculate that the nanoholes in the thin films seem to affect the magnetization reversal but do not have a significant impact on the current percolation paths. For the magnetotransport measurements we observe the $cos^{2}(\theta)$ dependency expected for the AMR effect. Depending on $m$, we find values of the relative AMR effect ranging from 4.4 to 9.5~$\%$. In particular, for the samples prepared with Ni:Fe precursors pulse ratio $m = 4, 5$ and 6, we register AMR$\%$ values in agreement with both what is expected for NiFe alloys in this compositional range \cite{McGuire} and the value measured on a reference thin film of evaporated commercial permalloy (dashed line). The process with $m=6$ results in the lowest spin-wave damping as will be discussed below. Here, we extract a relative AMR effect amounting to 5.6~$\%$. This value is larger than the one found for low-damping ALD-grown Ni which was 3.9~$\%$ \cite{Giordano2020}. We register an anomalous high AMR$\%$ value for the NiFe sample prepared with $m = 7$, and we attribute it to the different morphology and the larger number of holes in the thin film.
\begin{figure*}[h]
\centering
	\includegraphics[width=1\textwidth]{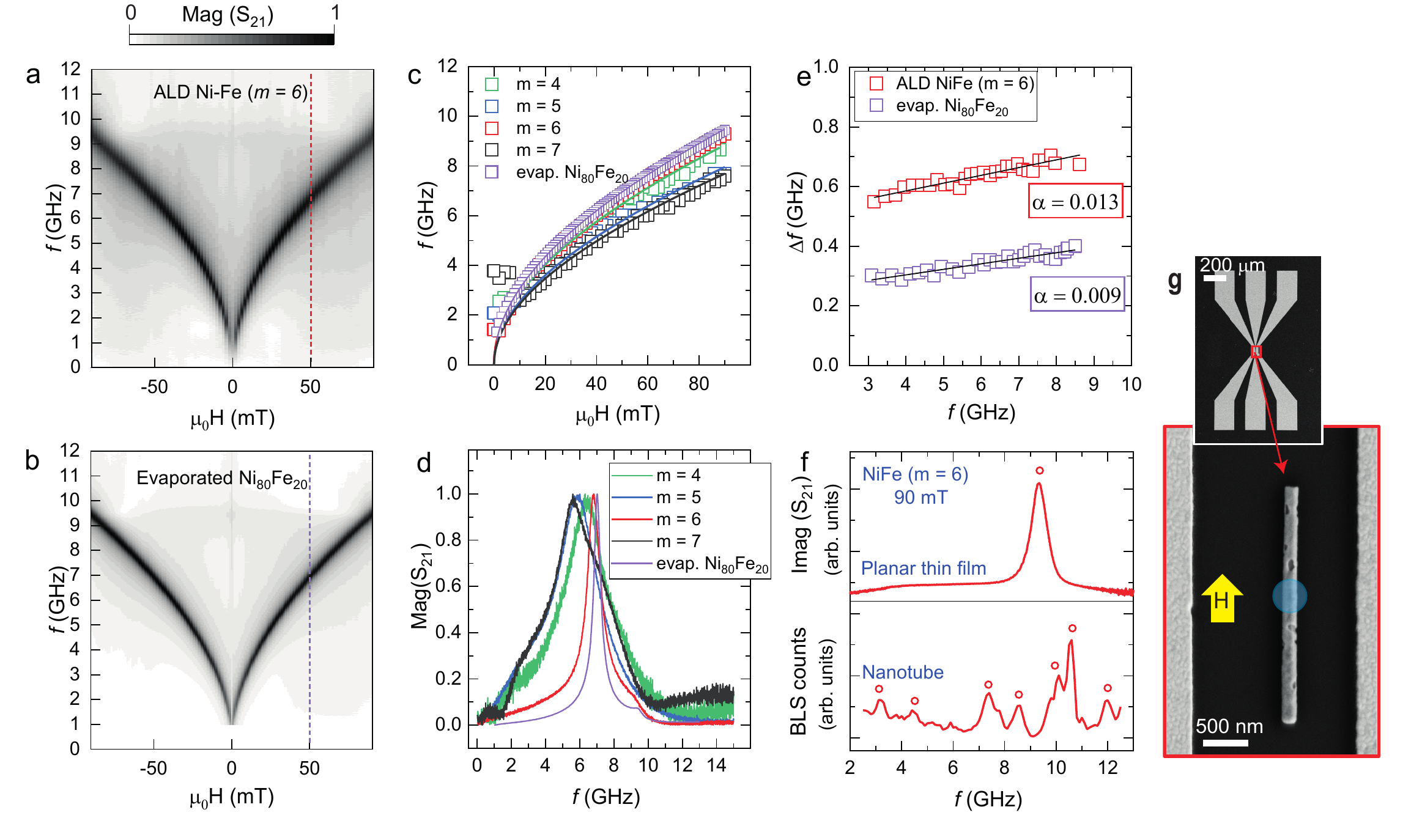}
	\caption{Magnitude of the scattering parameter $S_{21}$ measured at $T=300$~K in an in-plane magnetic field by VNA-FMR on (a) the NiFe thin film prepared by PEALD with $m = 6$ and (b) the e-beam evaporated thin film of \ce{Ni80Fe20} used as a reference. Black indicates large absorption. (c) Field dependent resonance frequencies (symbols) extracted from VAN-FMR spectra. The lines depict fits using the Kittel equation  \cite{Kittel1948} and are guides to the eye. (d) Individual spectra extracted at $+50$~mT and fitted by a Lorentzian function for the NiFe thin films prepared by PEALD using $m = 4, 5, 6$ and $7$ and the evaporated (evap.) \ce{Ni80Fe20} thin film. (e) $\Delta f$ reflecting the linewidth of the imaginary component of $S_{21}$ for the PEALD film prepared with $m = 6$ and the reference permalloy film plotted as a function of the resonance frequency. Each dataset is fitted by a linear function, whose slope is twice the damping parameter $\alpha$. (f) Imaginary component of the scattering parameter $S_{21}$ measured by VNA-FMR on the NiFe thin film prepared with $m=6$ (top) and $\mu$-BLS spectrum acquired on the central position of the NT prepared with the same process (bottom). The planar thin film is magnetized with an in-plane magnetic field of 90 mT, while the nanotube is magnetized along its long axis with a field of 90 mT. The resonance frequencies are indicated with red circles. (g) SEM micrographs of both the CPW employed to excite spin waves and the investigated NT lying parallel to the CPW's signal line. The blue circle indicates the position of the blue laser employed to detect the excited spin waves eigenmodes. The yellow arrow represents the direction of the in-plane magnetic field.}\label{Figure 7}
\end{figure*}

Now we report on the magnetization dynamics of the annealed NiFe samples, investigated by broadband ferromagnetic resonance (FMR) spectroscopy (Figure~\ref{Figure 7}).  Figure~\ref{Figure 7}a and b (Figure~S6) show the spectra of the evaporated commercial \ce{Ni80Fe20} thin film and of the PEALD-grown thin films prepared with $m = 6$ ($m = 4, 5$ and 7), respectively. The spectra were taken by inductive measurements on a coplanar wave guide (CPW) using a vector network analyzer (VNA) and reflected the magnitude of the scattering parameter $S_{21}$. The black branches in the color-coded spectra represent resonant absorption at different values of the in-plane magnetic field. In Figure~\ref{Figure 7}c we compare resonance frequencies for $H > 0$ extracted from spectra of the PEALD prepared NiFe thin films and the evaporated permalloy. In Figure~\ref{Figure 7}d we compare the individual spectra taken at the same field of 50 mT for thin films prepared with Ni:Fe pulse ratios $m = 4, 5, 6$ and $7$ and for the evaporated permalloy thin film used as a reference. The linewidths $\Delta f$, extracted from fitting a Lorentzian function to the spectra $S_{21}$, were employed to determine the damping parameter $\alpha$ of the materials (Methods). In Figure~\ref{Figure 7}e we report $\Delta f$ as a function of the resonance frequency for the PEALD sample prepared with $m = 6$ (upper curve) and the reference permalloy sample (lower curve). The curves are fitted by a linear function, whose slope is twice the damping parameter $\alpha$ \cite{Gurevich96}. The extracted values $\alpha$ are displayed next to the curves. The value $\alpha=0.013$ obtained for the ALD-grown thin film is larger than 0.009 detected on the reference NiFe film. The hole like features in the PEALD film most likely induce two-magnon scattering \cite{Tserkovnyak2005,Arias1999} and increase $\alpha$ beyond 0.009.

In Figure~\ref{Figure 7}f we show the imaginary component of the scattering parameter $S_{21}$ measured on a NiFe thin film prepared with $m=6$ by VNA-FMR and compare it with the spectrum obtained by microfocused Brillouin Light Scattering ($\mu$-BLS) on a nanotube prepared with the same PEALD process. The spectrum was measured at the central position of the NT as depicted in Figure~\ref{Figure 7}g. Both spectra were acquired at $\mu_{0}H$ = 90 mT. Resonance frequencies are indicated with open red circles. 
Figure~\ref{Figure 7}g shows a SEM micrograph of both the coplanar wave guide (CPW) employed to excite the spin precession in the individual NT and the NT parallel to the CPW's signal line. The blue circle indicates the position of the blue laser employed to detect the excited spin waves modes via $\mu$-BLS. The direction of the external applied field is indicated with a yellow arrow. The $\mu$-BLS spectrum of the permalloy NT is much richer than the ones reported for PEALD-prepared nickel nanotubes \cite{Giordano2020}. The series of resonance peaks indicate confined spin waves as previously reported for Ni NTs \cite{Giordano2020}. We attribute the larger number of resolved spin-wave resonances in Figure~\ref{Figure 7}f to the lower damping of NiFe compared to Ni. A detailed discussion about the nature of the additionally observed modes is beyond the scope of this paper.\\
Our results show that in the case of the Ni:Fe pulse ratio $m=6$ in the ALD deposition process we achieve field-dependent resonance frequencies (Figure~\ref{Figure 7}c) and a Gilbert damping parameter $\alpha$ (Figure~\ref{Figure 7}d) which are close to those values measured for a standard permalloy thin film of stoichiometry \ce{Ni80Fe20}. When studying the linewidths of field dependent resonant modes detected locally on the NT by means of $\mu$-BLS, we determined a Gilbert damping parameter of 0.01 (Figure~S7). The $\mu$-BLS technique allowed, in fact, to focus the laser on a defect-free region of the nanostructure. Thereby extrinsic contributions to the peak broadening were less dominant. The spin-wave spectroscopy confirmed the good quality of the ALD-grown permalloy and its reduced damping compared to ALD-grown Ni \cite{Giordano2020}, opening new possibilities for applications in 3D nanomagnonics.

\section{Methods}
\subsection{Plasma Enhanced Atomic Layer Deposition of NiFe and Subsequent Annealing}
Si (100) wafers were cleaved in pieces of about 2 cm x 2 cm and employed as substrates for the deposition of planar NiFe thin films. The Si substrates were cleaned with the following procedure: 15 min in \ce{H2O} : \ce{NH4OH} (28$\%$) : \ce{H2O2} (5:1:1) at 70$^\circ$C; 10 s in \ce{HF} (49$\%$) : \ce{H2O} (1:10) at room temperature; 15 min in \ce{H2O} : \ce{HCl} (37$\%$) : \ce{H2O2} (6:1:1) at 75 $^\circ$C.
GaAs NWs were grown on Si (111) substrates as fully described in references \citep{Matteini2015, Matteini16} and used as nanotemplates. The Si wafers were used both as bare substrates and coated with a 5 nm - thick layer of ALD alumina (\ce{Al2O3}) as previously described \cite{Giordano2020}. GaAs NWs were employed pre-coated with alumina. Nickel-Iron growth experiments were performed in a hot wall Beneq TFS 200 ALD reactor, operated at a pressure of 4-5 mbar, under a 100 sccm costant flow of ultrahigh purity nitrogen, used both as carrier and purge gas. We used nickelocene (\ce{NiCp2}), iron(III) tert-butoxide complex (\ce{Fe2(OtBu)6}) and water as precursors and reactant. They were stored in stainless steel containers at 80$^\circ$C, 100$^\circ$C and room temperature, respectively, to exploit their vapor pressure. The chamber temperature was set as 170$~^\circ$C. The plasma was generated in an RF parallel plate system and powered at 150 W. Pure hydrogen was supplied through the plasma head with a flow rate of 300 sccm. The PEALD sequence can be summarized as follows: [(\ce{NiCp2}/purge/\ce{H2O}/purge/\ce{H2} plasma/purge) $\times m$ + (\ce{Fe2(OtBu)6}/purge/\ce{H2O}/purge/\ce{H2} plasma/purge)] $\times$ n. The corresponding steps duration was the following: [(2s/4s/4s/8s/4s/8s) $\times$ m + (2s/4s/4s/8s/4s/8s)] $\times$ n  . NiFe thin films with thickness of 25 $\pm$ 3 nm were prepared setting $m = 4, 5, 6$ and 7 and using $n = 300, 250, 215 $ and 190, respectively, in order to keep the total number of precursors pulses $(m + 1) \times n$ equal to $\sim$1500, for each process. The thickness of the deposited thin films was measured by imaging the films in cross section by SEM. The growth rate was calculated by dividing the thickness by the total number of precursors pulses $(m + 1) \times n$. Thin films prepared on Si, Si / \ce{Al2O3} planar substrates and GaAs / \ce{Al2O3} NWs were thermally treated at 380$^\circ$C for 2h 30 min under forming gas \ce{N2} - \ce{H2} with flow 300 sccm. This temperature was reported to be sufficient to activate the reduction of iron oxide in the presence of nickel \cite{Espejo2016}.\\
Reference permalloy thin films were prepared in an electron-beam evaporator Leybold Optics LAB 600H, using a commercial target of \ce{Ni80Fe20}. A 50-nm-thick thin film was used as a reference for comparison in the paper, except for the measurements of resistivity and anisotropic magnetoresistance. Here we employed a reference \ce{Ni80Fe20} sample with a thickness of 20 nm comparable to the one of annealed NiFe samples prepared by ALD.

\subsection{Structural Characterization and Chemical Analysis of the Thin Films and Nanotubes}
We report the properties of the annealed thin films. Their morphologies and thicknesses were investigated by scanning electron microscopy (SEM) from Zeiss and Bruker and atomic force microscopy (AFM) from Bruker. Chemical analysis of thin films were performed by SEM combined with energy dispersive X-ray spectroscopy (SEM-EDS) on a Zeiss Merlin system (beam energy: 15 keV). X-ray diffraction spectra were recorded in the glancing incidence mode on a Malvern Panalytical(Empyrean model) diffractometer with incidence angle of 0.8$^\circ$. The morphologies of the annealed nanotubes was investigated by both SEM and transmission electron microscopy (TEM), chemical element distribution was examined by scanning transmission electron microscopy (STEM) combined with energy dispersive X-ray spectroscopy (STEM-EDS). The TEM and STEM experiments were carried out using an FEI Talos electron microscope operated at 200~kV. The thicknesses of the NTs were extracted from EDS elemental 2D maps of Ga and Ni using the software Velox, as previously discussed \cite{Giordano2020}. The diffraction pattern was simulated using the software JEMS, for the Ni:Fe crystal structure with atomic ratio 75:25.

\subsection{Investigation of Physical Properties of Thin Films}
The static magnetic properties of the thin films were assessed using a Microsense EZ-7 vibrating sample magnetometer (VSM) operated at room temperature. The saturation magnetization $M_{\rm s}$ values were determined from the VSM hysteresis by normalizing the magnetization M (emu) measured at 150 mT by the volume of the material V$_{m}$. For the calculation of V$_{m}$ the estimated volume of holes measured by AFM has been substracted. Resistivity measurements were performed with a KLA Tencor OmniMap RS75 four-point resistivity meter. Anisotropic magnetoresistance (AMR) measurements were carried out in the van der Pauw four-point configuration \cite{Kateb2019} in a custom-built set-up described previously \cite{Giordano2020}. AMR measurements were performed at room temperature applying a current of 10 $\mu$A and a static in-plane magnetic field of 80 mT applied at an angle varying from 0 to 360$^\circ$.
Broadband spectroscopy of the deposited thin films was performed using a vector network analyzer (VNA), sweeping the frequency and recording the ferromagnetic resonance (FMR) absorption spectra. The thin films were positioned on top of a CPW connected by microwave tips to the VNA. The 2-port VNA generates a microwave current providing an in-plane rf-magnetic field perpendicular to the long axis of the CPW. The frequency of the microwave magnetic field was swept from 10 MHz to 15 GHz. The microwave with a power of $-10~$dBm was applied at port 1 of the CPW in order to excite magnetization precession. The precession-induced voltage was detected at port 2 via the scattering parameter $S_{21}$ where the numbers 2 and 1 in the subscript denote the detection and excitation port. An external magnetic field $\mu_0H$ was swept from 90 mT to $-90~$mT along the CPW's long axis.
The field dependent frequency behavior was assessed by fitting the data by the Kittel equation \cite{Kittel1948,Gurevich96}. The Gilbert damping parameter $\alpha$ was determined by plotting the linewidth $\Delta f$ against the corresponding resonance frequencies. The curve was fitted by a linear function, whose slope is twice the damping parameter $\alpha$. The $\Delta f$ of the imaginary component of the scattering parameter $S_{21}$ was estimated dividing by $\sqrt{3}$ the linewidth of the magnitude component of $S_{21}$ ($\Delta$f Mag)\cite{Neusser2009}.
The metallic coplanar wave guide (CPW) for the investigation of the NT was fabricated by electron beam lithography and a following evaporation of 5 nm Ti/ 120 nm Au film. The signal line, having a width of 2.5 $\pm$ 0.1 $\mu$m was separated by gaps of 1.7 $\pm$ 0.1$\mu$m width from the ground lines. The CPW was fabricated around individual NTs placed parallel to the signal line and electrically bonded to a printed circuit board, which was connected to a signal generator (Anritsu MG3692C) applying a microwave current. The corresponding magnetic microwave field excited spin precession in the NT at a fixed frequency. Spin wave eigenmodes were detected via microfocused Brillouin light scattering ($\mu$-BLS) microscopy at room temperature \cite{Demidov2008IEEE}. A monochromatic blue laser with a wavelength of 473 nm and power of 0.5 mW was focused directly on top of the NT.

\section{Conclusions}
NiFe thin films with different Ni:Fe atomic ratios were prepared by alternating $m$ times the sequence for the deposition of nickel with a single sequence for the deposition of oxidized iron and a post-deposition annealing treatment in hydrogen. We achieved an optimized stoichiometry via the Ni:Fe pulse ratio $m=6$. By a further annealing step the resistivity of the optimized permalloy (Py) thin films was 28 $\mu\Omega$cm and the measured spin wave damping 0.013. A high relative AMR of 5.6~$\%$ was observed in the ALD-grown Py thin film with lowest spin wave damping. The high quality of the films allowed us to measure multiple resonant spin-wave eigenmodes in an individual Py nanotube. In the case of Py NTs the spectra were richer compared to the previously reported ALD-grown Ni NTs substantiating the lower damping of Py. The permalloy thin films and nanotubes thereby exhibited physical properties that make them promising for functional spintronic elements and magnonic applications in 3D device architectures.

\section*{Conflicts of interest}
There are no conflicts to declare.

\section*{Acknowledgements}
We thank SNF for funding our research via grants 163016, BSCGI0{\_}157705, NCCR QSIT and 197360. We thank Didier Bouvet, Bi Wen Hua and Ivica Zivkovic for the excellent experimental support.



\balance


\bibliography{library} 
\bibliographystyle{rsc} 


\section*{Supplementary material}

# Ni$_{80}$Fe$_{20}$ Nanotubes with Optimized Spintronic Functionalities Prepared by Atomic Layer Deposition

M. C. Giordano[1], S. Escobar Steinvall[2], S. Watanabe[1], A. Fontcuberta i Morral[2,3], and D. Grundler[1,4]

[1] Institute of Materials, Laboratory of Nanoscale Magnetic Materials and Magnonics, Ecole Polytechnique Federale de Lausanne, School of Engineering, 1015 Lausanne, Switzerland
[2] Institute of Materials, Laboratory of Semiconductor Materials, Ecole Polytechnique Federale de Lausanne, School of Engineering,1015 Lausanne, Switzerland
[3] Institute of Physics, School of Natural Sciences, Ecole Polytechnique Federale de Lausanne, 1015 Lausanne, Switzerland
[4] Institute of Microengineering, School of Engineering, Ecole Polytechnique Federale de Lausanne, 1015 Lausanne, Switzerland

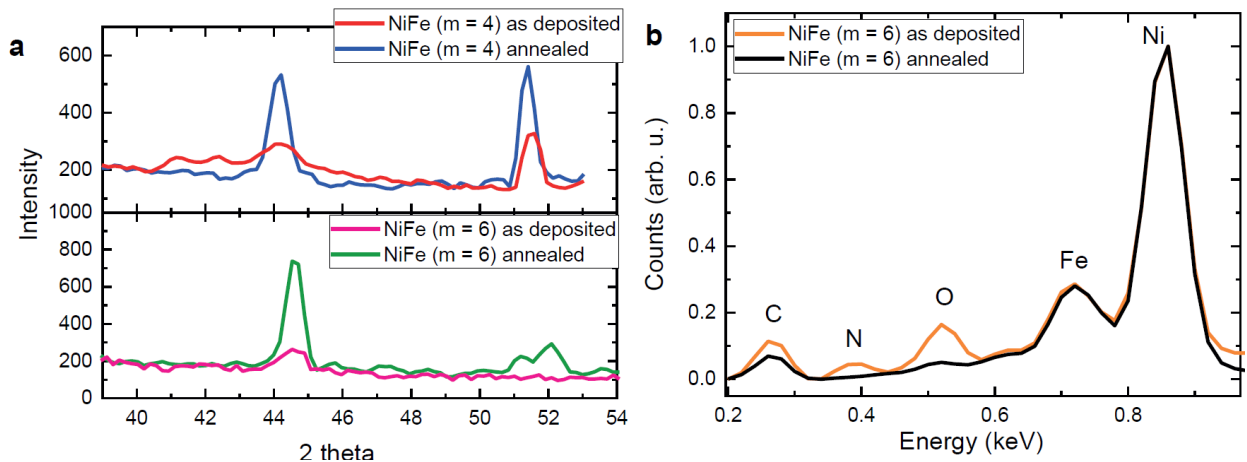

**Figure S1. (a)** X-Ray diffractograms of the samples prepared with $m = 4$ and $m = 6$ in their as deposited and annealed state. **(b)** Energy dispersive spectroscopy of the NiFe sample prepared with $m = 6$ as deposited and annealed in the energy range 0.2-1 keV. The annealing treatment was performed at 380 °C.

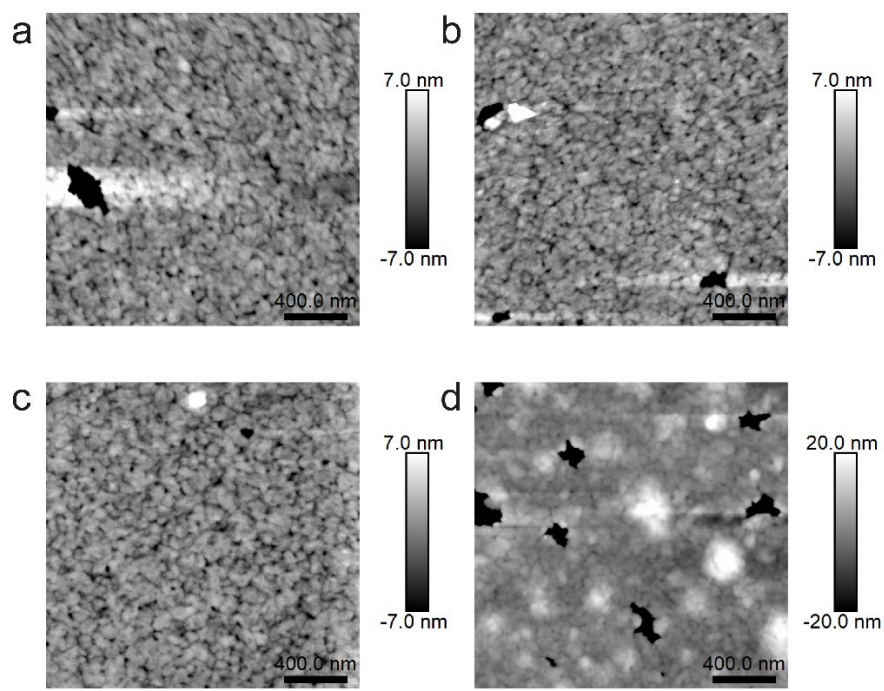

**Figure S2.** Atomic force microscopy performed on a 2 μm × 2 μm area of samples prepared with (a) *m* = 4, (b) *m* = 5, (c) *m* = 6, and (d) *m* = 7 on bare silicon substrates, after the annealing treatment at 380 °C.

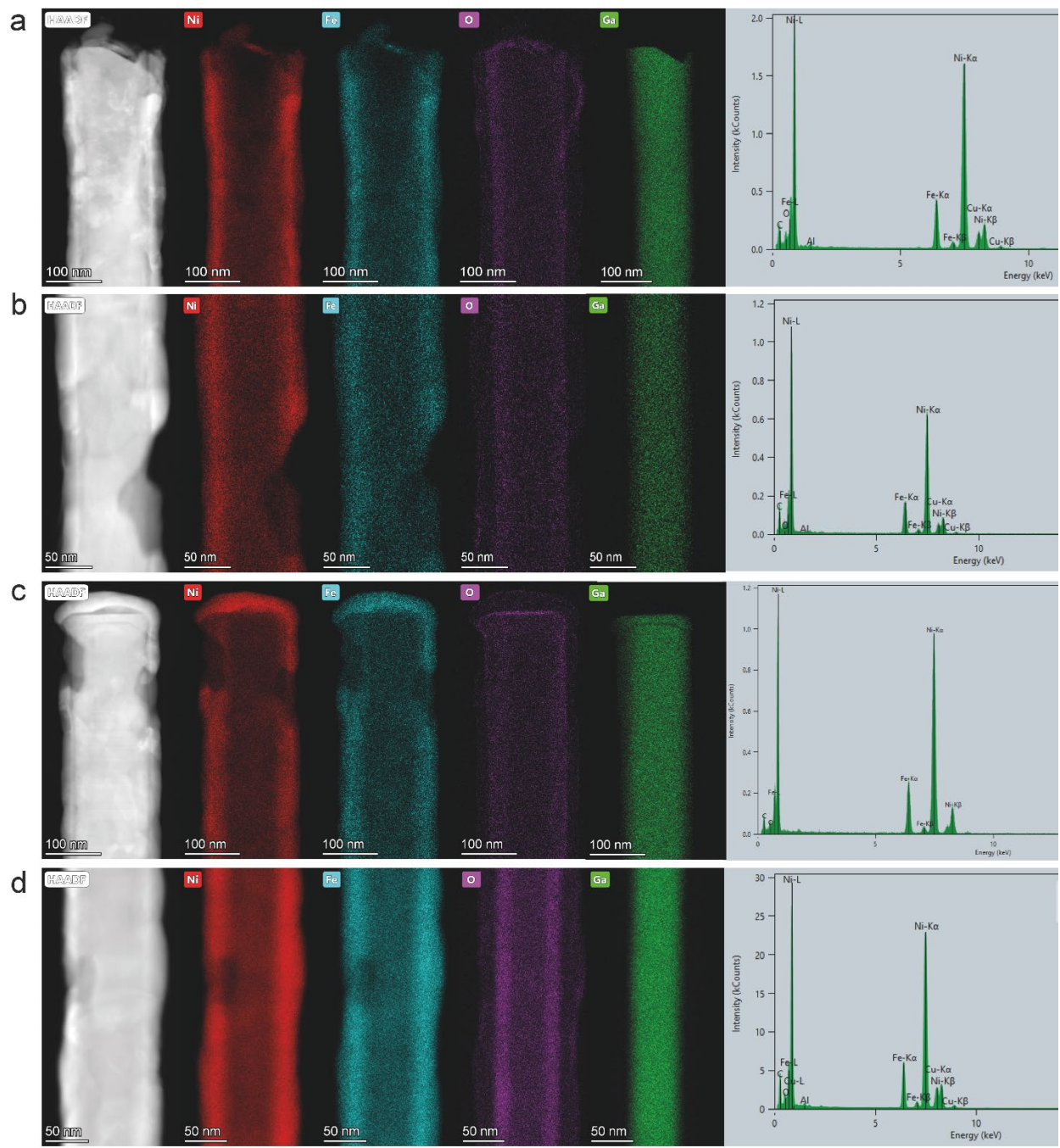

**Figure S3.** HADDAF images and elemental maps of NiFe nanotube samples (a-d) NT_S1-4, prepared with Ni:Fe pulse ratio *m* = 6, annealed at 380 ˚C, and corresponding energy dispersive spectroscopy analysis of the external shell. The composition and NiFe shell thickness values measured for the four nanotubes are summarized in Table S1.

**Table S1.** Thickness and compositional values of NiFe NTs prepared by ALD with Ni:Fe pulse ratio $m$ = 6. The values were extrapolated from STEM-EDX analysis.

| Sample | Thickness far from holes (nm) | Ni (at%) in NiFe ratio | Fe (at%) in NiFe ratio | O (at%) in the shell |
|---|---|---|---|---|
| NT1 (main text) | 21.1 | 80.2 | 19.8 | 3.5 |
| NT_S1 | 23.3 | 80.7 | 19.3 | 3.4 |
| NT_S2 | 20.2 | 80.5 | 19.5 | 3.6 |
| NT_S3 | 23.3 | 80.7 | 19.3 | 3.5 |
| NT_S4 | 22.1 | 80.8 | 19.2 | 3.5 |
| **Average** | 21.6 ± 1.2 | 80.4 ± 0.3 | 19.6 ± 0.3 | 3.5 ± 0.1 |

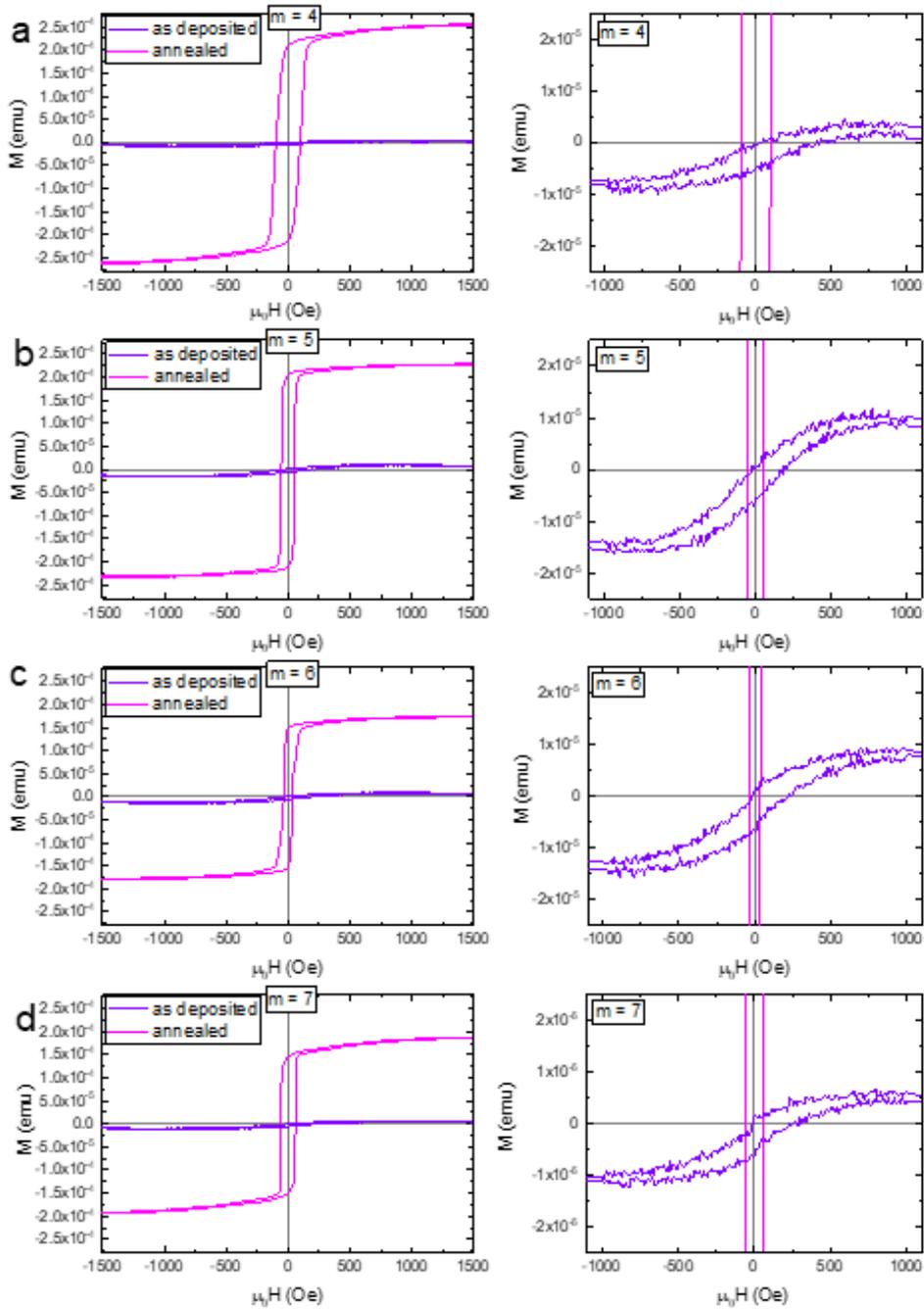

**Figure S4.** Hysteresis curves measured with a VSM, at room temperature, of as deposited (purple curve) and annealed (magenta curve) NiFe thin films prepared by ALD, using Ni:Fe pulse ratio (a) m = 4, (b) m = 5 , (c) m = 6 and (d) m = 7. The hysteresis curves were acquired with an applied field oriented at 0˚ degrees in the plane of the sample. On the right panel the same curves are magnified.

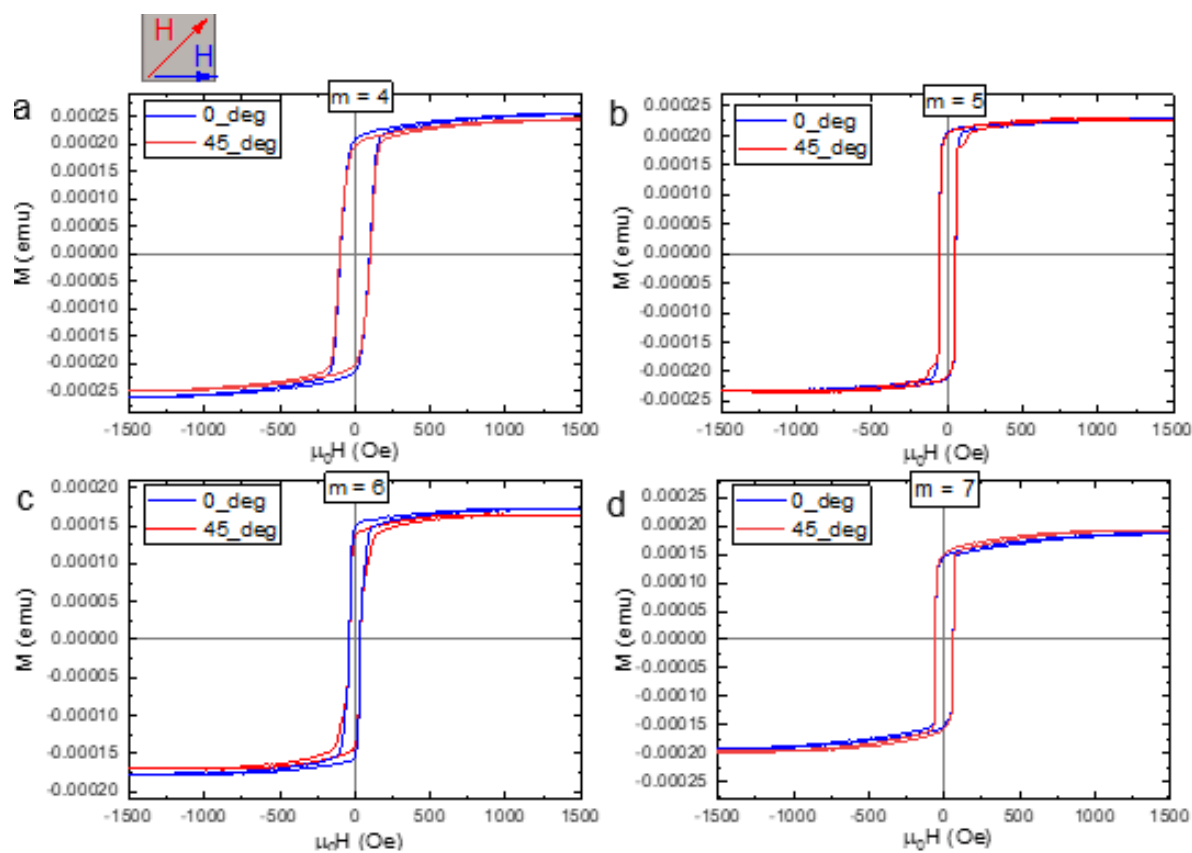

**Figure S5.** Hysteresis curves measured with a VSM at room temperature of annealed NiFe thin films prepared by ALD, using Ni:Fe pulse ratio (a) $m$ = 4, (b) $m$ = 5 , (c) $m$ = 6 and (d) m = 7. The blue (red) curves were acquired with an applied field oriented at 0 (45) degrees in the plane of the sample.

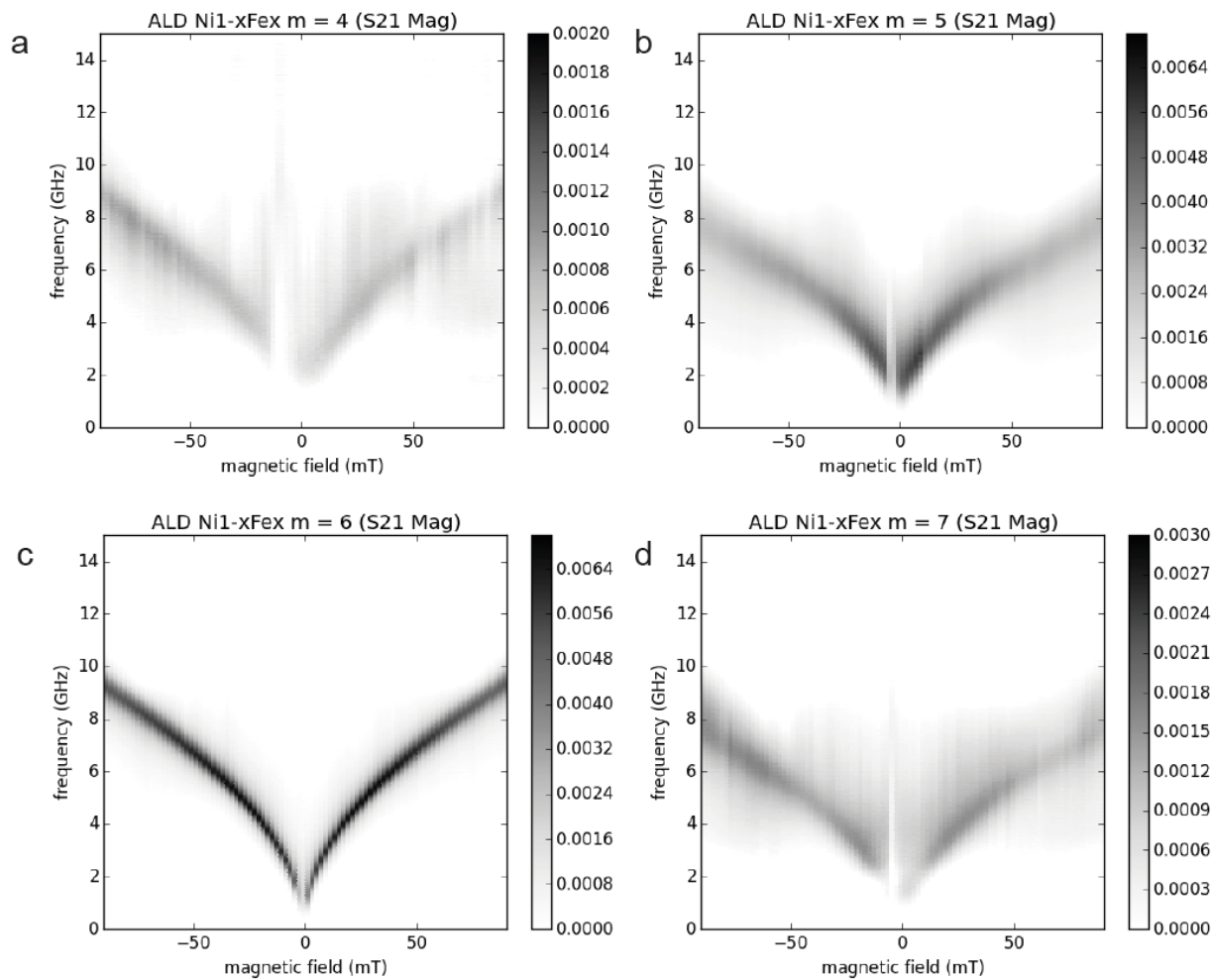

**Figure S6.** VNA-FMR measurements of annealed NiFe thin films prepared by ALD, using Ni:Fe pulse ratio (a) $m = 4$, (b) $m = 5$, (c) $m = 6$ and (d) $m = 7$. The resonance measurements were acquired while sweeping the magnetic field from 90 to -90 mT in the plane of the sample.

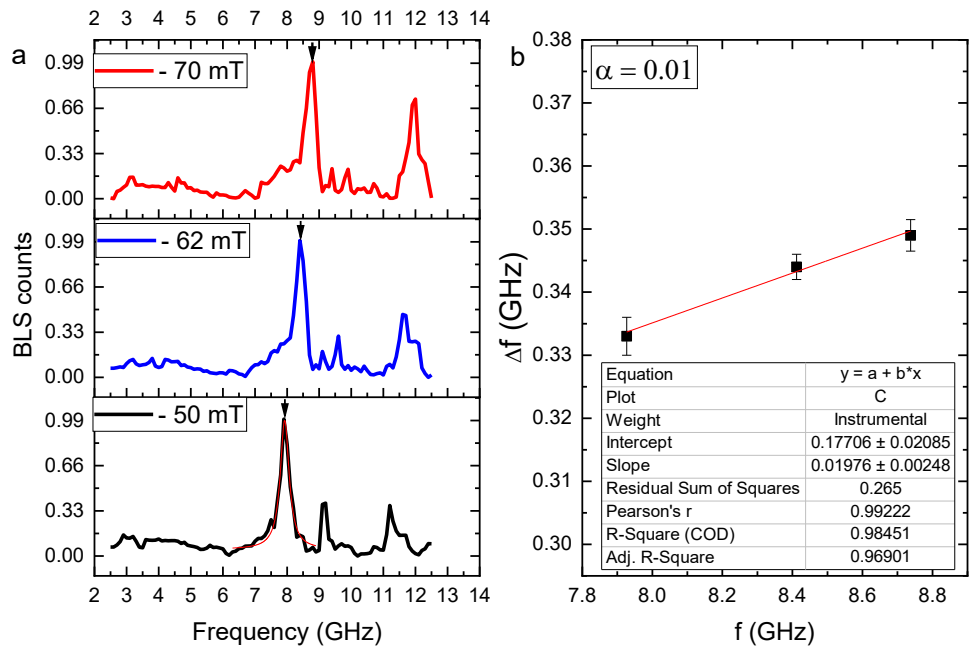

**Figure S7.** (a) BLS spectra detected, at different applied static magnetic fields, on an individual permalloy NT prepared by PEALD using Ni:Fe pulse ratio *m* = 6. Black arrows indicate the peak (eigenmode) whose linewidth was used to assess the NT damping parameter. (b) Linewidth Δf plotted as function of the resonance frequency of the mode selected. The data are fitted with a linear function whose slope is twice the Gilbert damping parameter α, that is found to be 0.01 consistent with the VNA-FMR data.


\end{document}